\documentclass{emulateapj}

\def\approx{$\sim$}
\def\persqcm{$\rm cm^{-2}$}
\def\persqpc{$\rm pc^{-2}$}
\def\percc{$\rm cm^{-3}$}
\def\h2{$\rm H_2$}
\def\error{$\pm$}
\def\e#1{$\times 10^{#1}$}
\def\tenup#1{10$^{#1}$}
\def\asec{\arcsec}
\def\amin{\arcmin}
\def\deg{\arcdeg}
\def\thr{$^h$}
\def\tmin{$^m$}
\def\tsec{$^s$}

\def\kms{km~s$^{-1}$}

\def\solmass{$\rm M_{\sun}$}

\newcommand{\beamsz}[2]{${#1}'' \times {#2}''$}

\newcommand{\jykms}{$\rm Jy\,km\,s^{-1}$}
\newcommand{\mjb}{$\rm mJy\,b^{-1}$}

\begin{document}

\title{The Aptly Named Phoenix Dwarf Galaxy}
\author{Lisa M.\ Young,}
\affil{Physics Department, New Mexico Tech, 801 Leroy Place
Socorro, NM 87801}
\email{lyoung@physics.nmt.edu}
\author{Evan D.\ Skillman, Daniel R.\ Weisz, }
\affil{Astronomy Department, University of Minnesota, 116 Church St. SE,
Minneapolis, MN 55455}
\email{skillman@astro.umn.edu, dweisz@astro.umn.edu}
\and
\author{Andrew E.\ Dolphin}
\affil{Department of Astronomy/Steward Observatory, 933 N.\ Cherry Ave.,
University of Arizona, Tucson, AZ 85721-0065; Raytheon Company}
\email{adolphin@raytheon.com}

\slugcomment{ApJ, accepted 27 December 2006}

\begin{abstract}
The Local Group galaxy Phoenix has the properties of a dwarf spheroidal
galaxy, but an adjacent HI cloud has been recently found to be at the same
radial velocity as the stars.  The proximity suggests that this cloud
is associated with the most recent ($\le 100$ Myr) star formation in Phoenix.
We have obtained relatively
high sensitivity and high resolution HI imaging with the VLA with
the goal of distinguishing between different processes for displacing 
the gas from the galaxy.  Due to the outer curvature of the HI cloud,
it appears that expulsion from the galaxy by winds from supernovae
is more likely than ram-pressure stripping.  The isolation of the 
galaxy makes tidal stripping highly unlikely.  
Using a star formation history constructed from HST imaging, we construct a
simple kinematic model which implies that the HI cloud is still
gravitationally bound to the galaxy. 
Gas which is expelled from the centers of dwarf galaxies but which remains gravitationally
bound may explain the episodic star formation observed in several dwarfs.
In the specific case of Phoenix, there
may be future star formation in this currently dSph-like galaxy.

\end{abstract}

\keywords{
galaxies: individual (Phoenix dwarf) ---
galaxies: dwarf ---
galaxies: ISM ---
galaxies: kinematics and dynamics ---
galaxies: evolution 
}

\section{Introduction}

The Local Group dwarf galaxies are an intriguing puzzle.  Most
of them can be divided into two different families: the
dwarf spheroidals (which show no HI gas, no current star formation, and
no sign of rotational support) and the dwarf irregular galaxies (which
are HI rich, and show obvious signs of 
recent star formation and a significant degree of rotational support).
In the last decade it has become clear that the dwarf spheroidals show a
large range in star formation histories--- some of them are dominated
by old ($\sim$10 Gyr) stellar populations, but many have large components of
intermediate age ($\sim$5 Gyr) and even some young ($\le$1 Gyr) stars
\citep{mateo98,grebel01,dolphin06}. 
Stellar abundance patterns also show that in many cases the process of
recycling interstellar gas into new stars must have continued over multiple
Gyr \citep[e.g.,][]{ikuta02,venn04,fenner06}.
Thus, 5 billion years ago, many of the galaxies which we call dwarf spheroidals
must have been significantly more gas-rich than they are now and they
may have looked very much like today's dwarf irregulars.
What mechanism or mechanisms are responsible for transforming the dwarf
spheroidals into their current state?

The fate of the neutral gas in the smallest galaxies is relevant to
cosmological problems of current interest, such as the chemical enrichment
history of the early universe.   Additionally, it may offer insight into
the ``dwarf galaxy number crisis"-- the prediction from standard cosmologies
that hundreds of unidentified dark matter halos should inhabit the Local
Group \citep{moore98, klypin99}.
A long favored theory holds that bursts of star formation can remove the
ISM from a dwarf galaxy because of the small gravitational potential depth
\citep[e.g.,][]{L74,V86,DS86}.
However, recent hydrodynamical simulations of bursts in dwarf spheroidal
systems imply that removing all of the ISM may be difficult to accomplish in
practice \citep{MF99,marcolini06}.
This theory may also require some delicate fine-tuning in order 
to explain a galaxy like the Carina dwarf, which 
shows evidence of three well isolated episodes of
star formation \citep{saha86,mighell90,smecker-hane94,mighell97}.
If the star formation activity is energetic enough to clear gas from the
galaxy, then either fresh gas was re-accreted between episodes
or the clearing is only partial.

Ram pressure stripping presents a second viable theory, and it has 
the advantage of naturally explaining the morphology-density relationship
observed for the Local Group dwarfs
\citep[e.g.,][]{E74, LF83, K85, V94a, V94b}.
Recently, \citet{mayer01a,mayer01b,mayer06} have
suggested that ram pressure plus ``tidal stirring'' (the combined effects on a satellite galaxy
during passage through perigalacticon) can convert a dwarf irregular into
a dwarf spheroidal by inducing a burst of star formation while simultaneously
heating the stellar disk and pulling off the ISM in a tidal stream.
These processes are most important, though, on orbits with pericenters less
than 50 kpc.

High resolution and high sensitivity HI observations provide 
a probe to examine which
processes are responsible for removing the gas from the galaxy.
For ram pressure stripping, one expects
the gas to be displaced in the opposite direction to the galaxy's motion.
One might also expect that on the side of the cloud facing the optical galaxy
there would be a stronger column density gradient giving a bow shock
appearance similar to Ho II \citep{bureau02} and the LMC \citep{deboer98}.
For blow-away, naively, one expects gas ejected
in more than one direction and possibly the presence of a bow
shock pointed away from the galaxy.  
For tidal stirring, one expects an elongated tidal feature.

The Phoenix dwarf offers a unique opportunity in the study of the
mechanisms that can transform a gas-rich dIrr into a gas-poor dSph.
Its stellar population is primarily old, but some star formation activity
has continued until about 100 Myr ago \citep{ortolani88,MGA}. 
There is also an HI cloud which is associated with the galaxy
\citep{YL97,stgermain,gallart01} but displaced from the stellar body, and this
arrangement suggests that whatever processes are responsible for gas
removal might be taking place right now in Phoenix.
The other nearby dwarf
galaxies have either lost their gas long ago, so that no signs of the
removal process remain, or they still retain their gas, so that no signs
of the removal process are visible yet.

The first map of the $-23$ \kms\ HI cloud now believed to be associated 
with the Phoenix dwarf was presented by \citet{YL97} at \beamsz{133}{102} 
resolution.  Subsequent mosaic imaging by \citet{stgermain} had comparable
sensitivity and resolution (\beamsz{180}{180}) but much better recovery of
large scale ($\gtrsim$ 10\arcmin) HI structures in the region. 
At the time, the radial velocity of the Phoenix dwarf was not known, but 
more recent
velocities from stellar spectra \citep{gallart01, irwin02, held02} showed
an excellent match with the HI radial velocity, securing the association.
Comparisons of the properties of the $-23$ \kms\ cloud with the other 
nearby gas clouds provided evidence that the $-23$ \kms\ cloud is indeed 
physically associated with the Phoenix dwarf and that the others are 
almost certainly foreground emission.
We now present very deep VLA integrations on the $-23$ \kms\ cloud which
give a factor of three better sensitivity at improved resolution
(\beamsz{71}{71}).
We use these high quality images to study the detailed structure of the
cloud and to seek clues, as described above, to the physical mechanisms
responsible for the evolution of the interstellar medium in the Phoenix
dwarf.

The distance to the Phoenix dwarf has been very well constrained by 
observations of its tip of the red giant branch stars since
the pioneering work of \citet{VDK}, who first derived a distance modulus
of 23.1 $\pm$ 0.1.  Subsequent studies have all been consistent with 
this value as \citet{MGA} derived 23.0 $\pm$ 0.1, 
\citet{held99} derived 23.04 $\pm$ 0.07 (and a value of 23.21 $\pm$ 0.08
based on the luminosity of the horizontal branch stars), and
\citet{holtz00} derived a value of 23.11 with an uncertainty of 
``several percent.''  As described in \S 4, we have re-analyzed imaging
from the HST archive, and find a distance modulus of 23.11 $\pm$ 0.05
(or, roughly 420 $\pm$ 10 kpc).  We will use this distance throughout
the paper.

\section{Observations and Data Reduction}

The first interferometric HI observations of the Phoenix dwarf consisted of 
2.0 hours with the National Radio Astronomy Observatory's Very Large Array
\footnote{The 
National Radio Astronomy Observatory
is a facility of the National Science Foundation operated under cooperative
agreement by Associated Universities, Inc.} 
in the DnC configuration on 1996 May 18 and 4.0 hours in the CnB configuration
on 1996 Jan 23.  The results of those data are published in \citet{YL97}.

Additional observations in 2003 and 2004 were, in total, four times longer
than the 1996 data.
The Phoenix dwarf was observed for a total of 12.0 hours in the DnC
configuration on 2003 Jan 24, 25, and 27 and for another 12.0 hours in the
CnB configuration on 2004 Jan 23, 24, and 25.
All data were obtained with a total bandwidth of 1.56 MHz, giving 256
channels of 6.1 kHz (1.3 km s$^{-1}$) each.
The velocity range covered is +188.7 \kms\ to $-74.2$ \kms\ (heliocentric,
in the optical definition).
Integration times were 60 seconds on source in the DnC configuration and 20
seconds on source in CnB.
The nearby point source J0155-408 was observed every 40 minutes as a phase
reference; the bandpass and flux calibration data were obtained from the
source J0137+331.

All data calibration and image formation were done using standard calibration
tasks in the AIPS package.
Initial imaging revealed which channel ranges were free of HI line emission.
Continuum emission was subtracted directly from the raw visibility data by making first
order fits to the line-free channels.
All datasets were then combined.  For a low resolution map (described in
greater detail below), the datasets were combined with no change to 
their nominal weights.  For a higher resolution map the data weights in
the 2004 CnB observations were multiplied up by a factor of two.  This
upweighting partly offsets the lower weights resulting from the shorter
integration times in the CnB array so that
after final gridding in the visibility plane, both DnC and CnB configurations
contribute roughly equally to the final image.

The primary imaging difficulty for VLA imaging of the Phoenix dwarf 
is in achieving a nearly round synthesized beam for a source of such
low declination.
The beam is elongated in the north-south direction even for data from the DnC
configuration, so, in effect, the north-south baselines from the CnB
configuration are used in combination with the east-west baselines from the
DnC configuration to create a synthesized beam which is nearly round.
A low resolution cube was made using the ``natural" weighting
scheme (data weights do not depend on the local density of samples) applied
with a Gaussian taper to downweight long baselines.
The resulting beam had a fitted FWHM of 130\asec~$\times$~116\asec.  
The dirty image was cleaned down to a residual level of approximately 1.0
times the rms noise fluctuations, and clean components were restored with a
circular beam of FWHM 130\asec.
A higher resolution cube was made using 
Briggs's robust weighting scheme \citep{SIRA2}, as implemented in the 
AIPS task IMAGR with a robustness parameter 2, plus an additional Gaussian
taper to downweight baselines extended in the east-west direction.
The synthesized beam for this image had a fitted FWHM of
71\asec~$\times$~52\asec.  It was cleaned in the same manner as the low
resolution cube and restored with a round beam of FWHM 71\asec.
Offline Hanning smoothing produced channels of width 2.6 \kms\ for both
cubes.
Additional imaging details and final sensitivities are given in Table
\ref{obstable}.

Integrated intensity (zeroth moment) maps were made 
by the masking method:
the deconvolved image cube was smoothed along both spatial and velocity
axes, and the smoothed cube was clipped at about 1.0$\sigma$ (the noise
level before smoothing) in absolute value.
The clipped version of the cube was used as a mask to define a
three-dimensional volume in which the emission is integrated over velocity.
Integrating the ``moment zero" image again over the spatial directions then
gives the total HI flux.

\begin{deluxetable}{lcccc}
\tablewidth{0pt}
\tablecaption{Image Parameters
\label{obstable}}
\tablehead{
\colhead{} & \colhead{high res} & \colhead{low res}  \\
}
\startdata
uvtapers (k$\lambda$) & 3, 50 & 1, 3 \\
robustness & 2 & 5 \\
``original" beam FWHM (\asec) & 71 $\times$ 52 & 130 $\times$ 116 \\
``original" beam PA ($^\circ$) & 14  &  57 \\
restoring beam FWHM (\asec) &  71 $\times$ 71 & 130 $\times$ 130 \\
Linear resolution (pc) & 140 & 260 \\
rms noise (\mjb) &  1.2 & 1.9 \\
N(HI) sensitivity (cm$^{-2}$) & 2.0$\times 10^{18}$ & 9.6$\times 10^{17}$ \\

\enddata
\tablecomments{
Column density sensitivities are 3$\sigma$ in one channel.}
\end{deluxetable}

The low resolution HI cube contains a total of 2.95 \error 0.05 \jykms\ of
HI emission in the vicinity of the Phoenix dwarf.
The uncertainty in this value is indicative of the variability that results
from slightly different sum regions.
The higher resolution HI cube contains a total of 2.8 \error 0.1 \jykms,
which is entirely consistent with the flux in the low resolution cube. Both
values
are just a bit higher than the 2.6 \jykms\ measured by \citet{YL97} in
their lower sensitivity data cubes.
Accounting for uncertainty in the absolute flux calibration,
we count a total flux of 2.95 \error\ 0.10 \jykms\ from the Phoenix dwarf.
In comparison, \citet{stgermain} quote 4.0 \jykms\ in this cloud, so 
the new VLA maps have recovered at least 75\% of the HI flux.
At 420 kpc, this flux corresponds to 1.2\e{5} \solmass\ of HI.

\section{The Spatial and Kinematic Structure in the Phoenix HI Cloud}

As in previous observations of the region, Galactic emission is detected in
the velocity range +16 to $-2$ \kms.
The southern clouds noted by \citet{YL97} and \citet{stgermain} are also 
evident in the velocity range +44 to +75 \kms.
However, since these clouds are not believed to be associated with the
Phoenix dwarf we do not discuss them further.

\begin{figure*}
\centerline{\includegraphics[width=18cm,clip]{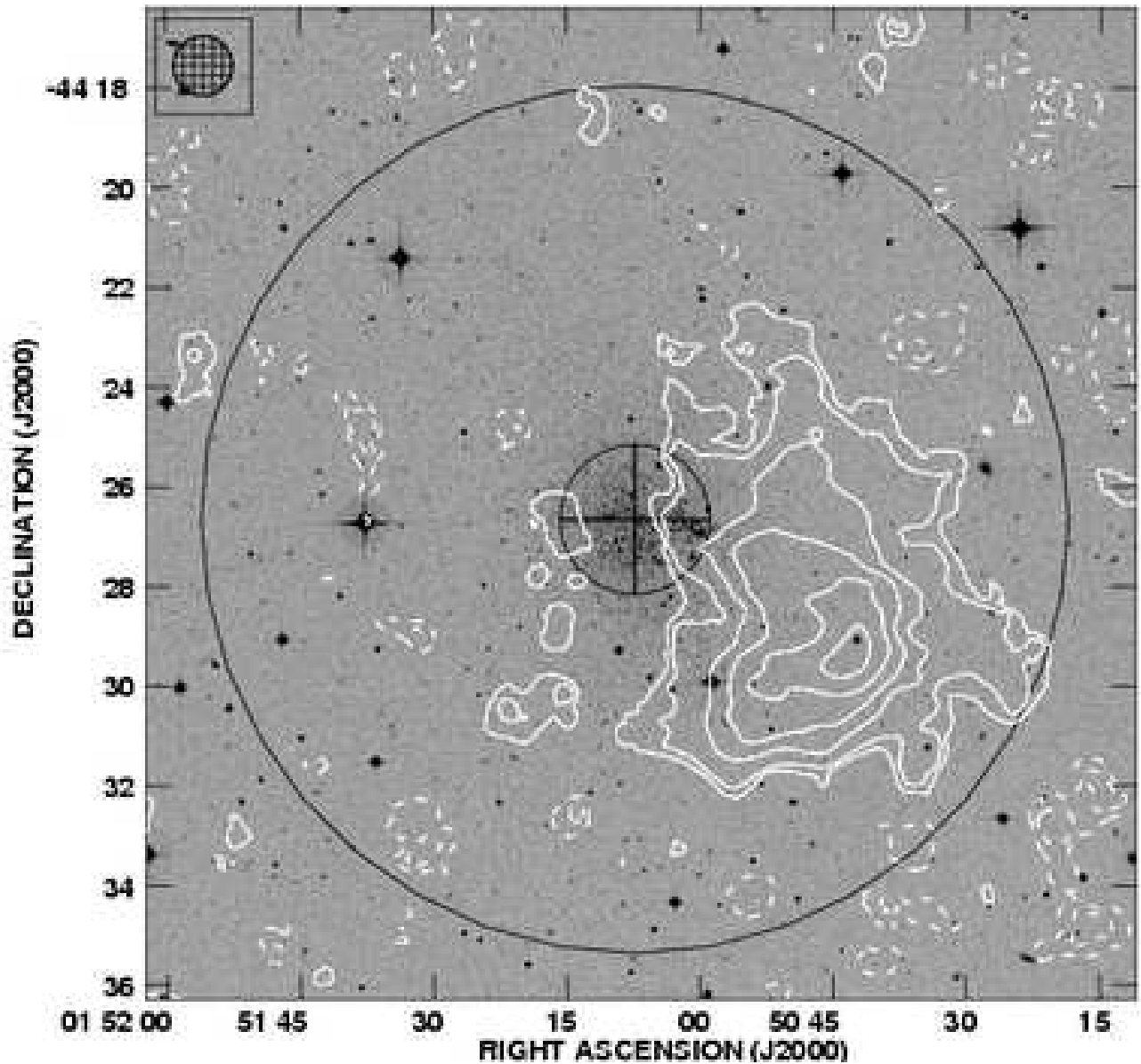}}
\caption{Integrated HI intensity at 71\asec\ resolution overlaid on 
the optical image of Phoenix from the DSS.  Contours are
($-2, -1, -0.5$, 0.5, 1, 2, 3, 4, and 5) \e{19}\ \persqcm.
The small hashed circle in the top left corner indicates the resolution of
the HI image.
The crossed circle marks the bright part of the stellar distribution and is
intended to guide the eye for a comparison with Figure \ref{channels}.
Note that the stellar distribution of Phoenix
has been traced out to a radius of 8.7\arcmin\ \citep{VDK}, as shown by the
large circle, so that the HI
observed here is within the projected area of the Phoenix galaxy.
\label{mom0}}
\end{figure*}

Our new images confirm the general impressions gained from earlier, lower
resolution and lower sensitivity data but provide important new details on
the internal structure of the gas cloud associated with Phoenix.
For example, Figure \ref{mom0} shows the integrated HI intensity of the
$-23$ \kms\ cloud at 71\asec\ resolution.
At column densities \approx \tenup{19} \persqcm\ the angular extent of the
cloud is 
roughly 8\amin $\times$ 8\amin, similar to that noted by \citet{stgermain}.
However, the curved shape of the
cloud is much more obvious here than in the low resolution images.
The new data also reveal that the ``center of curvature" of the cloud is
{\it not} coincident with the optical center of the galaxy; rather, it is
some 2\amin\ to 3\amin\ (260 to 390 pc) south of the optical center.
However, the northern-most and southern-most extensions of the HI cloud do
lie perfectly symmetrically above and below the declination of the center 
of the optical galaxy. 
Note that the small apparent optical diameter of Phoenix is a bit deceptive.
The stellar distribution has been traced out to a radius of 
8.7\arcmin\ \citep{VDK}, so that the HI
observed here is within the projected area of the Phoenix galaxy.

\begin{figure*}
\centerline{\includegraphics[width=18cm]{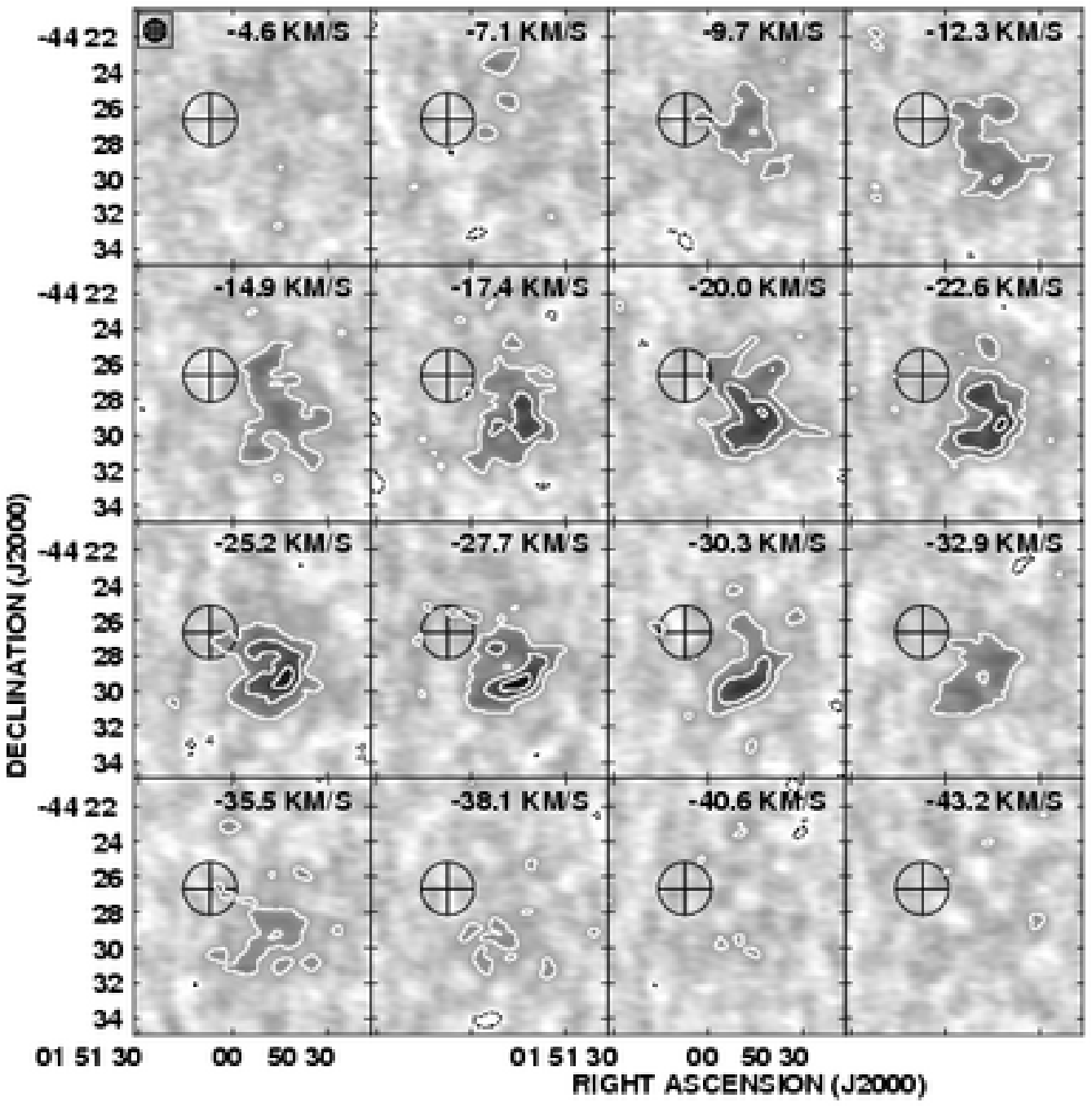}}
\caption{Channel maps at 71\asec\ resolution.  Contours are
$-3$, 3, 6, and 9 times the rms noise (1.2 \mjb).
The resolution is indicated in the top left corner of the first panel, and
the crossed circle shows the position and approximate size of the stellar
body visible in the DSS image of Figure \ref{mom0}. 
\label{channels}}
\end{figure*}

Figure \ref{channels} shows individual channel maps at 71\asec\ resolution; 
the curved shape of the HI in many of the channels and a gradual north-south 
velocity gradient are clearly evident.  The peak brightness temperature 
(1.5 K) is not in the center of the cloud but in its southwest quadrant 
near 01\thr\ 50\tmin\ 44\tsec,
$-44$\deg\ 29\amin\ 30\asec.  The peak column density 
(5.4\e{19} \persqcm, or 0.6 \solmass~\persqpc\ including He) is also here, so
that the steepest radial gradients in column density are on the side opposite
the stellar body.
The peak brightness temperature at 71\asec\ resolution is only modestly
higher than the brightness temperature at \beamsz{133}{102} resolution
(1.4 K)
even though the beam area is smaller by a factor of nearly 3, which suggests
that the primary structures in the HI cloud are now resolved.

\begin{figure}
\includegraphics*[scale=0.5]{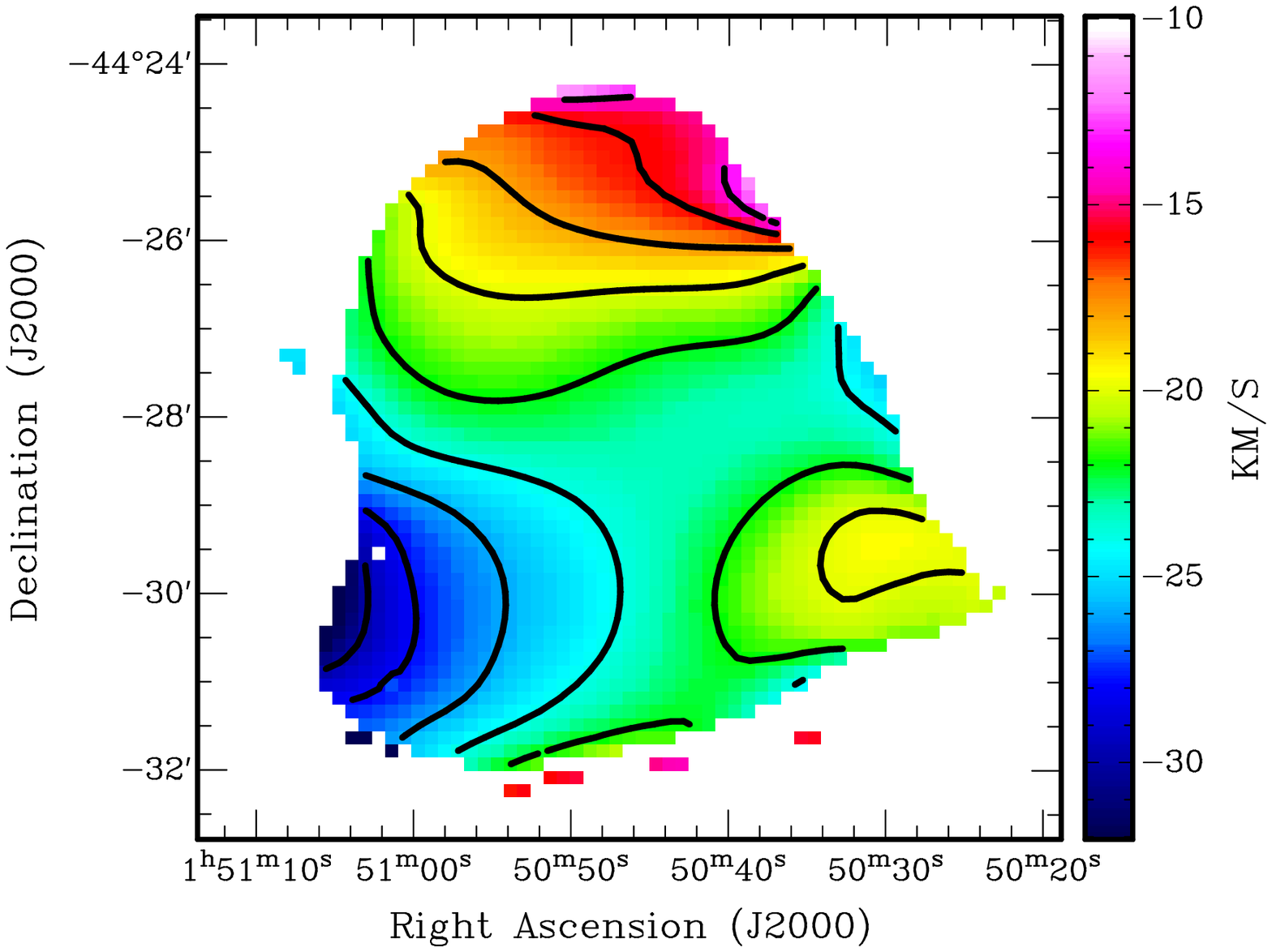}
\caption{Velocity field, derived from fitting Gaussian line profiles to the
image at 130\asec\ resolution.  Contours range from $-30$ to $-14$ \kms\ in
steps of 2 \kms.
\label{velfield}}
\end{figure}

Figure \ref{velfield} shows the velocity field of the HI
emission; it also confirms general features noted in the channel maps and
in the velocity field of \citet{stgermain}.  The HI cloud shows a velocity
gradient primarily in the north-south direction from $-18$ \kms\ on the
north side towards $-28$ \kms\ in the south.  Beyond $-23$ \kms\ the
isovelocity contours bend towards the southeast, so that the most negative
velocities are actually in the southeast corner of the cloud (east of
01\thr~51\tmin\ and south of $-44$\deg~29\amin).  The integrated intensity
image (Figure \ref{mom0}) also shows a protuberance on the west side of the
HI cloud (01\thr~50\tmin~30\tsec, $-44$\deg~29.5\amin) which departs from the
velocity trend noted above.  Figure \ref{velfield} shows that the protuberance
has velocities of $-20$ \kms\ and higher; it is also visible in channel
images at $-12$, $-15$, and $-20$ \kms.  Even though this feature has quite
low column density, only 1\e{19} \persqcm\ at 71\asec\ resolution, it is
clearly detected in both the ATCA observations of \citet{stgermain} and in
our VLA data, so it is undoubtedly real.

\begin{figure}
\includegraphics*[scale=0.5]{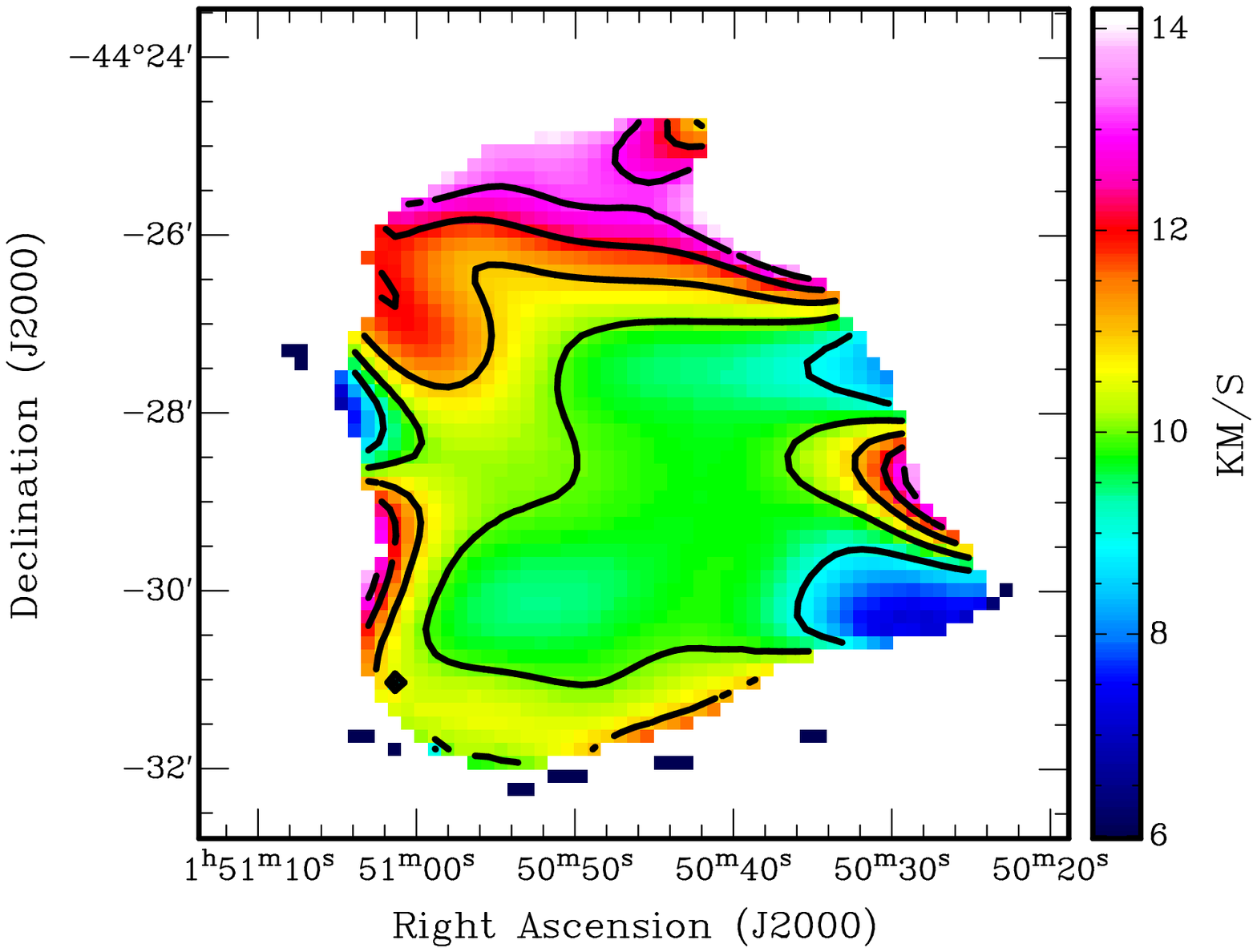}
\caption{Velocity dispersion at 130\asec\ resolution.  Contours range from 9
\kms\ to 13 \kms\ in steps of 1 \kms.
Most of the HI can be characterized as having 
a velocity dispersion between $\sim$9  and 11 \kms.
A few pixels in which the formal uncertainty in the dispersion is greater than 2.5
\kms\ are not shown.
\label{dispersions}}
\end{figure}

Fits of a Gaussian line profile to the data in velocities $-2.0$ \kms\ to
$-38.1$ \kms\ give the velocity dispersions shown in Figure
\ref{dispersions}.  Those HI dispersions range from 8 \kms\ to 14 \kms, and,
to first order, most of the HI is consistent with a value of 10 \kms .
There is a trend for larger dispersions to be found in the northern part of the
cloud, and, aside from noisy regions at the edge of the galaxy, the
dispersions are somewhat larger on the east side (\approx~11 \kms)
than on the west (\approx~9 \kms).  
This trend, with larger dispersions on the side closer to the optical
galaxy, is consistent with the pattern expected from an expanding shell.
Looking through the outside edge of the shell, the expansion velocity would be
tangential to the line of sight and would not contribute to the line width;
closer to the center of the expanding shell (on the north and east sides of the
HI cloud, in this case) the expansion velocity would have some observable component
and would broaden the line.
However, if the HI cloud is an expanding shell its expansion velocity is
only a few \kms\ and it is not possible to distinguish the expansion
signature in position-velocity slices dominated by the overall N-S velocity
gradient.

Neither our present data nor the results of \citet{stgermain} give any
hints for the presence of additional HI emission offset in any other direction 
from the optical galaxy, down to column density limits of approximately 
\tenup{18} \persqcm\ at
130\asec\ resolution (40 times smaller than the observed column densities).
The significance of this non-detection lies mainly in the question of
the importance of ram pressure as a gas removal mechanism.  HI gas offset 
in multiple directions would almost certainly rule out ram pressure
stripping as a credible explanation; however, we find no such gas. 

In considering the possible effects of ram pressure on the gas in Phoenix,
though, the morphology of the HI cloud is particularly interesting when
compared to the morphology of HI in the dwarf galaxy Holmberg II
\citep{bureau02}.
The low resolution map of Ho~II clearly shows smooth, convex, and
compressed HI contours on the southeast side of the galaxy, with more
irregular and fluffy low column density material extending to larger radii
on the north and west sides.
The cometary appearance of the HI in Ho~II is clear evidence for the
effects of ram pressure as the galaxy moves through the M81 intra-group
medium.
If ram pressure were responsible for displacing the HI from the center of
Phoenix,
we would expect to find a steeper HI column density gradient on the side
towards the stellar body.  We might also expect the cloud to have a convex
side towards the stellar body.  Neither of those is true, and thus the
comparison with Ho~II suggests that ram pressure is {\it not} responsible
for the displacement of the HI in Phoenix.

Quantitatively, it is also unlikely that the density of the halo gas is high
enough to effect any ram pressure stripping at Phoenix.  In the usual manner
we can estimate the necessary density from the relation \citep[e.g.][]{GG72}
$P_{ram} \sim  \rho_{halo} v_{gal}^2 \sim GM_{gal} \Sigma_{gas}/R^2,$
where $v_{gal}$ is the velocity of the galaxy through the halo gas, $M_{gal}$
is the mass of the galaxy interior to some radius $R$, and $\Sigma_{gas}$ is
the column density of the atomic gas there.  Making the most conservative
estimates for the HI properties of Phoenix we consider the outer edge of the
HI cloud, which has surface
density 0.1 \solmass~\persqpc\ (including helium) at 590 pc from the galaxy center.  We take
Phoenix's mass to be 3.3\e{7}~\solmass\ \citep{mateo98} and its velocity
about the Milky Way approximately 220 \kms.  Stripping then requires the
density of the halo gas to be $\rho_{halo} \gtrsim$~3\e{-5} \percc.
Stripping the gas of the highest observed column densities from the center of
Phoenix requires at least an order of magnitude larger $\rho_{halo}$.
\citet{murali00} has estimated that the density of the Milky Way halo must be
$\lesssim 10^{-5}$ \percc\ at the location of the Magellanic Stream, however,
and Phoenix is some eight times further from the Milky Way than that.
Of course ram pressure stripping is easier if the mass of Phoenix has been
overestimated, but by analogy with \citet{marcolini06}, if the mass of
Phoenix is very much smaller than we have assumed here it could not have 
retained its ISM long enough to support an extended star formation history.

Tidal stirring is another process which can be responsible for the removal
of gas from a galaxy.  However, since this process is most relevant at
distances less than 50 kpc from the Milky Way \citep{mayer06}, 
the relative isolation of Phoenix makes this 
scenario unlikely.  The radial velocity of $-$23 km s$^{-1}$ corresponds to a 
galactocentric radial velocity of $-$112 km s$^{-1}$ using the formulae presented
in \citet{RC3}.  At 112 \kms\ it has been at least 3 Gyr since Phoenix was
close enough to the Milky Way for significant tidal interaction.  
If Phoenix is bound to the Milky Way, the semi-major axis of its orbit must
be at least 210 kpc, giving an orbital period of 7.6 Gyr for a Milky Way mass
of 1.4\e{12} \solmass\ \citep{mateo98}.  Clearly, interactions with the Milky
Way may have influenced Phoenix several Gyr ago but not more recently than
that.

\section{The Recent Star Formation History of the Phoenix Dwarf}\label{sfh}

In an early ground-based optical study of Phoenix, \citet{ortolani88}
noted the presence of a small ($\sim$ 10$^4$ M$_{\odot}$)
population of young ($\sim$ 10$^8$ yr)
blue stars.\footnote{The abstract of the 
paper states 10$^7$ yr, but the text assigns an age of 10$^8$ yr.}
From deeper photometry, \citet{MGA} (MGA) 
studied the distribution of these blue stars and found evidence of an
east-to-west progression across the face of Phoenix.  MGA estimated
the age of these stars to be 100 Myr, and, from this, the relative
position of the HI cloud detected by \citet{YL97}, and the
assumption that the HI cloud was associated with the recent star
formation, estimated an outward velocity for the HI cloud of 
$\sim$ 7 km s$^{-1}$.
At the time, the optical radial velocity of Phoenix was not known.

\citet{gallart01} obtained velocities for 31 Phoenix red giants
and determined an optical radial velocity of $-$52 km s$^{-1}$.  The
$\sim$30 km s$^{-1}$ difference between the optical radial velocity
and the HI radial velocity required a re-calculation of the estimated
HI cloud velocity.  The radial velocity difference implied a
trajectory nearly along the line of sight, because placing the HI cloud back
on top of the stars 100 Myr ago means the tangential velocity of the cloud is
much less than 30 \kms.  The resulting galaxy-HI
cloud separation is 3.5 kpc.  At this velocity and separation, the HI
cloud is not bound to Phoenix (the mass estimate falls short by
roughly one order of magnitude).

However, \citet{irwin02} obtained a new radial velocity of
$-$13 $\pm$ 9 km s$^{-1}$, which is in excellent agreement with the HI
velocity of $-$23 km s$^{-1}$.  \citet{held02} also reported agreement
between the optical and HI radial velocities within 2 -- 3 km
s$^{-1}$.  Thus, the original calculation by MGA would appear to be
more relevant than that of \citet{gallart01} concerning the gravitational status of the HI cloud.

\begin{figure}
\includegraphics[scale=0.4]{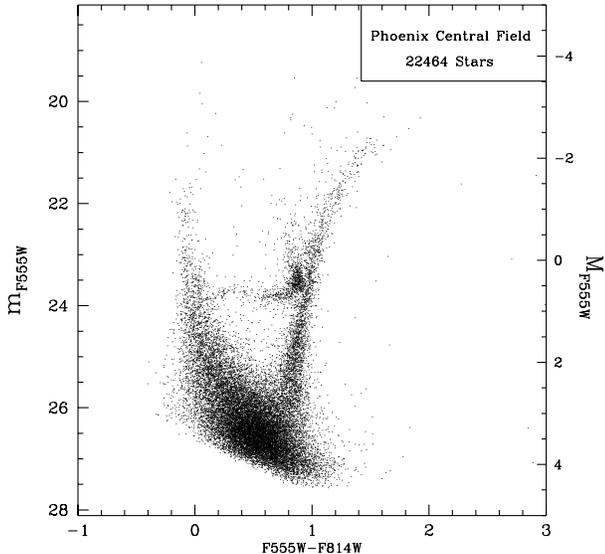}
\caption{The V versus V-I color magnitude diagram from the HST WFPC2 observations
of the central field of Phoenix (see text).  The photometry comes from the 
Local Group Stellar Populations Archive (Holtzman et al.\ 2006).
The right axis shows the absolute magnitude scale corresponding to the derived
distance modulus of 23.11 $\pm$ 0.05.
Note the truncation in the main sequence at an apparent magnitude of $\sim$ 21.5.
This provides an accurate determination of the age of the last star formation in 
Phoenix.
\label{CMD}}
\end{figure}

We have revisited the recent star formation history of the Phoenix 
dwarf in the following way.  The Hubble Space
Telescope WFPC2 has obtained relatively deep optical imaging of
Phoenix in two different programs.  \citet{holtz00}
report observations of a central field in Phoenix.  They derive
a star formation history for this field, and note that star formation
proceeded in Phoenix up until 100 Myr ago, in agreement with
\citet{ortolani88} and \citet{MGA}.
Deeper HST WFPC2 observations of a central
field which is almost coincident with that of Holtzman et al.\  were
obtained by Aparicio (HST-GO-8706) in 2001.  We have retrieved the
photometry from these two Hubble Space Telescope observations of
Phoenix from the HST/WFPC2 Local Group Stellar Photometry Archive
\citep{holtz06}.  For reference, the V,V--I color magnitude diagram 
for the deeper integration is shown in Figure \ref{CMD}.

\begin{figure}
\centerline{\includegraphics[scale=0.45]{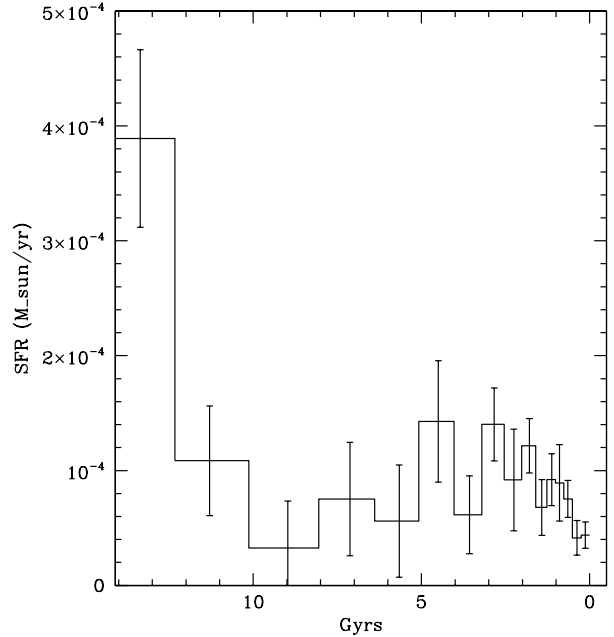}}
\caption{The global star formation history of the Phoenix dwarf galaxy
derived from HST observations of a central field (see text).
This star formation history is in generally good agreement with
the central field star formation history derived by 
\citet{holtz00}.
\label{GSFH}}
\end{figure}

Using the programs of \citet{dolphin02}, we have derived the star 
formation histories for both Phoenix central field HST observations.  
In Figure \ref{GSFH} we show the total star formation history for the 
central field derived from the deeper integration.  This star formation
history agrees well with that derived from the slightly shallower integration 
and the analysis of those observations presented by \citet{holtz00}.
There is evidence for a large amount of star formation at the earliest
ages, a relative decrease at intermediate ages ($\sim$ 10 -- 5 Gyr ago),
and then increased star formation activity in the last 5 Gyr which decreases
as the present time is approached.

\begin{figure}
\centerline{\includegraphics[scale=0.45]{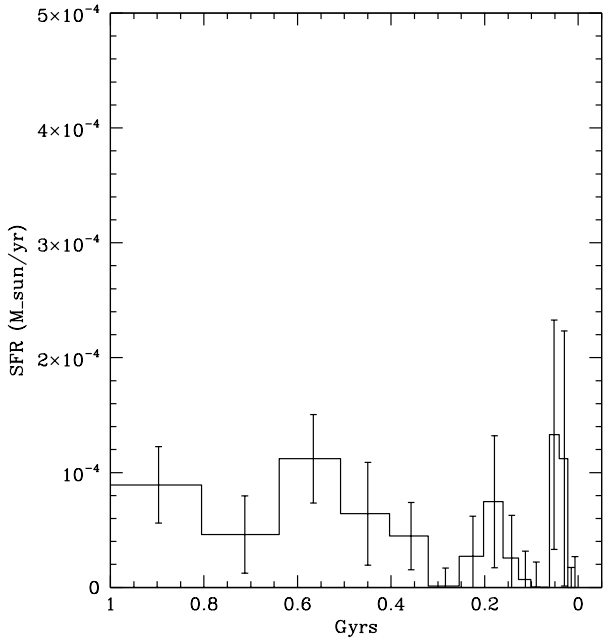}}
\caption{The recent star formation history of the Phoenix dwarf galaxy
derived from HST observations of a central field (see text).
The most recent episode of star formation has an age of roughly 
50 Myr with no evidence of star formation since then.
\label{RSFH}}
\end{figure}

In Figure \ref{RSFH} we show the star formation history for the 
last Gyr derived from the deeper integration at the higher time resolution
afforded by the observations.  The age of the last
star formation event appears to be about 50 Myr ago with a duration of
roughly 20 Myr.  A similar analysis of the shallower central field
observation gives a similar result.  
The age of 50 Myr for the most recent episode of star formation is 
slightly younger than the previous reported values of $\sim$100 Myr ago.
However, this is the first detailed analysis of this event from HST
photometry, and Figure \ref{CMD} shows that there is a clear truncation
in the main sequence stars at an apparent V magnitude of $\sim$ 21.5,
so the age estimate should be secure.  Thus, it is likely that 50 Myr 
is a better estimate than 100 Myr for the most recent star formation 
event.  Note that this most recent star 
formation episode is comparable in strength to the earlier episode at 
$\sim$200 Myr ago.  (Because linear time is plotted on the horizontal axis of
Figure \ref{RSFH}, the 
total star formation for any time is simply the area under the curve 
for the appropriate time bin.) Also note that the ``gaps'' in star formation
at $\sim$300 Myr and $\sim$100 Myr ago are of similar duration 
($\sim$50 Myr).  

Using the age of 50 Myr for the most recent star formation event,
we can calculate simple estimates of the HI cloud velocity and
determine its gravitational status.
The angular separation between the peak column density of the HI cloud
and the optical center of the galaxy is 289\arcsec . At a distance of 
420 kpc, this corresponds to a projected distance of 590 pc.
Under the assumptions that the HI cloud is 590 pc from the galaxy 
and was expelled by the most
recent episode of enhanced star formation, we derive a velocity of
$\sim$12 \kms.  An upper bound on the velocity and separation of the
HI cloud can be obtained by including the observed radial velocity
difference between HI and the stellar body (10 \kms); the recession speed of
the HI cloud could be as large as 16 \kms\ and its true separation as large
as 780 pc.

Given Phoenix's total mass of 
3.3\e{7} M$_{\odot}$ \citep{mateo98}, the escape velocity for the HI
cloud at the distance of 590 pc is 21 \kms.
At 780 pc from Phoenix, the escape velocity is 19 \kms.
The HI cloud's velocity is 3--9 \kms\ lower than the escape velocity, so
we find that the cloud is still gravitationally bound.  It is
likely that the HI will fall back and may serve as the raw
material for future episodes of star formation. 
However, we also note that there
is an observed velocity difference of 12 \kms\ from one side of the
cloud to the other, meaning that some portion of the cloud may be unbound. 
These conclusions are contradictory to those of \citet{gallart01}, but 
this is purely a result of the previous authors' use of a relatively
large radial velocity offset between the HI and the stars. 

For the observed HI mass, multiplied up by a factor of 1.3 to account for
helium, and the recession velocities discussed above, we find that the total
mechanical energy in the HI cloud is 2.3\e{50} -- 3.9\e{50} ergs.
Integrating over the star
formation history, we calculate a total mass of $\sim$5\e{3} \solmass\ 
of new stars formed in the recent star formation
event. For a standard IMF, one expects more than 10 stars with masses
greater than 10 M$_{\odot}$ formed in that event, so, even at low efficiency, there
is sufficient energy available from the supernovae associated with the
recent star formation to propel the HI away from the center of the
galaxy.

Finally, there remains the question of why neutral gas was only ejected to
one side of Phoenix.  Naively, one expects gas blown away by a burst of 
star formation to be ejected in all directions unless the energy source was
initially off-center with respect to the gas.  Could one off-center supernova
be responsible for the asymmetric gas displacement in Phoenix?  
Assuming a canonical total energy for a supernova of \tenup{51} erg implies
an energy transfer efficiency in the range of 20\%--40\%.
Note, however, that this efficiency does not take geometric effects into
account --- if the blast energy were injected symmetrically, much less than
half of the \tenup{51} erg would be available to accelerate a cloud off to
one side of the supernova.
Thus, our estimates are inconsistent with energy input from a single typical
supernova.
The idea also neglects the effects of the other $\sim$ 10
supernovae that likely occurred.  However, the combined effects of those
supernovae could be responsible for the single-sided expulsion.
Based on a comparison
between the distributions of the brightest main sequence stars and the 
red supergiants, \citet{MGA} found evidence
for star formation propagating across Phoenix from east to west.
The star formation wave may act 
as a plow to sweep up the gas and push it to one side of the galaxy.

We have investigated the HST observations to see if the detailed star
formation history supports this interpretation of single-sided expulsion.
The accuracy of the photometry of the main sequence stars in 
the HST photometry is substantially better than that of the ground based
observation used by \citet{MGA}, so it is possible for us to examine
the spatial distribution of main sequence stars as a function of luminosity
(or upper age limit).  In agreement with \citet{MGA}, we find the majority 
of the brightest main sequence stars fall in a band stretching across 
the middle of Phoenix.  The brightest of these (stars with indicative ages
of 25 - 75 Myr) are clustered in the association on the west side of the 
galaxy.  Stars with older ages are more centrally concentrated.  We also
derived separate star formation histories for the eastern side of the 
galaxy and the western side of the galaxy.  The western side shows evidence
for strong episodes of star formation at 40 and 180 Myr ago preceded
by star formation $\sim$ 400 Myr ago.  The eastern side shows evidence of 
relatively constant star formation at lower levels from 50 to 250 Myr ago 
and also star formation
at $\sim$ 450 Myr ago.  In sum, the data are consistent with a general
trend for the star formation propagating from east to west over the last 
200 Myr, but the pattern is not a clear-cut, monotonic one (i.e., both
sides of the galaxy participated in the star formation episode 200 Myr ago). 
These data are not inconsistent with the HI location on the west side of
the galaxy.

\section{Discussion}

The fact that the HI cloud appears to be bound to Phoenix is
intriguing.  From the asymmetrical shape of the HI cloud and the relative
isolation of Phoenix, it appears
that ram pressure
stripping is not a likely cause of its displacement.  Thus,
supernova winds appear the most likely explanation for the HI cloud to
be driven away from the center of the galaxy.  However, if the HI
cloud is bound, then $\sim$ 10$^5$ M$_{\odot}$ of HI can fall back into
the galaxy, perhaps triggering a future episode of star formation.

Indeed, stellar population work has already suggested that this
kind of a scenario (temporary HI expulsion, fall-back, and rejuvenated star
formation) may have happened in the Carina dSph.
Early on, \citet{mould82} noted that the presence of blue
stars in the Carina dwarf indicated that Carina and possibly other dSphs
were not single age systems.  Further evidence for a complicated
star formation history for Carina gathered 
\citep{saha86,mighell90,smecker-hane94,mighell97}
until the present day
picture of Carina consists of three episodes of star formation at roughly
old, intermediate (7 Gyr) and young (3 Gyr) epochs 
\citep{hurley-keller98,rizzi03,monelli03}.
Two major
questions which have always plagued studies of the star formation history
of the Carina dSph are: (1) what caused the large gaps between the episodes 
of star formation, and (2) what removed the neutral gas from Carina such 
that there is no longer any star formation at all?

On the second question, the final clearing of gas from Carina and indeed the
other dwarf spheroidals, Phoenix does not provide a direct answer because its
HI is apparently still bound.
The hydrodynamical simulations of \citet{marcolini06} and others also show
that it is difficult for star formation activity, by itself, to
blow out all of the ISM unless the dark matter halos are significantly
smaller than expected and/or the hot gas's radiative energy losses much
smaller.  The simulations do tend to lift a significant component of the
galaxy's cold gas up away from the stellar body in the
aftermath (30 Myr -- 50 Myr) of a burst, as we observe
in Phoenix.  But if star formation cannot completely unbind the ISM from a
dwarf spheroidal, it is commonly assumed that some external agent such as
tidal forces or ram pressure stripping must unbind the gas.  This idea is
certainly consistent with the fact that the most gas-poor dwarf spheroidals
such as Draco and Ursa Minor tend to be closer to the Milky Way or to M31
than Phoenix is.  A recent proper motion estimate for the Ursa Minor dwarf
spheroidal implies that its perigalacticon is $\lesssim 50$ kpc
\citep{piatek05}, carrying
it well into the region where interactions with the Milky Way can clean it of
its gas \citep{mayer06}.   \citet{piatek05} have made the interesting
comment, however, that perigalacticon distance cannot be the only factor
influencing the evolution of a galaxy as the Carina and Ursa Minor dwarfs
have similar orbits but quite different star formation histories.

With regard to the first question posed above, the origin of the gaps in the
star formation activity of Carina, 
the Phoenix dwarf may be providing us with some answers.  Here we posit
that a relatively modest amount of star formation has lifted the neutral gas
up out of the center of the potential well.
However, most of the gas is not
moving in excess of the escape velocity and is destined to return to the
center.
Again, from the separations and velocities discussed in \S \ref{sfh}
we can calculate the total energy (gravitational plus mechanical) of the HI
cloud, the semi-major axis of its orbit and its orbital period.
The estimated orbital periods are in the range 140 Myr to 660 Myr, so it
seems reasonable that this mechanism could produce gaps in the star
formation history on timescales of a few hundred Myr (Figure \ref{RSFH}).
On the 660 Myr orbit the semi-major axis is 1.2 kpc; thus,
the HI cloud could reach angular separations several times larger than the
current value.

Given the present distribution of the HI gas, it seems unlikely that it
will all fall back in at once, so there may be some additional time delay
as the gas collects before initiating the next episode of star formation.
Since only $\sim$ 10$^3$ M$_{\odot}$ of stars were formed in the episode
that dislocated the gas, we may expect a similar amount of star formation  
when the $\sim$ 10$^5$ M$_{\odot}$ of HI gas
returns to the galaxy's center.
Thus, the Phoenix dwarf may be aptly named.

\section{Summary}

We have presented new VLA HI observations of the Local Group galaxy Phoenix.
The galaxy has the properties of a dwarf spheroidal galaxy, with the
exception that an HI cloud is located at a similar velocity and within the
optical radius of the stellar distribution.  The close association of
the HI cloud with the galaxy suggests that this cloud may be the remnants
of gas responsible for the most recent star formation in Phoenix which 
took place $\le 100$ Myr ago.  

Our new high sensitivity and high resolution VLA HI imaging reveals that 
the HI distribution has a crescent-like shape with the center of 
curvature in the general direction of the galaxy.  This indicates that ram 
pressure stripping is unlikely to be responsible for the displacement of the gas
from the galaxy.  Additionally, the isolation of the galaxy makes tidal 
stripping highly unlikely.  

This leaves expulsion from the galaxy center by winds from supernovae
as the most likely explanation.
Using a star formation history constructed from HST imaging, we construct a
simple kinematic model which implies that the HI cloud is still
gravitationally bound to the galaxy.  Much of the atomic gas can be expected
to collect again in the center of the galaxy on timescales of a few hundred
Myr. 
Thus, there may be future star formation in this dSph-like galaxy.

\acknowledgments

Support for this work was provided by NASA through grant AR-9251 from the
Space Telescope Science Institute, which is operated by AURA, Inc., under
NASA contract NAS5-26555.
LY thanks New Mexico Tech for a sabbatical leave and appreciates the
hospitality of the University of Oxford sub-department of Astrophysics, where
part of this work was done.  EDS is grateful for partial support from NASA
LTSARP grant No.\ NAG5-9221 and from the University of Minnesota.  DRW is
grateful for support from a Penrose Fellowship.  
We also thank the referee for helpful comments.
This research has made use of NASA's Astrophysics Data System Bibliographic
Services and the NASA/IPAC Extragalactic Database (NED), which is operated by
the Jet Propulsion Laboratory, California Institute of Technology, under
contract with the National Aeronautics and Space Administration.

\end{document}